\documentclass[twocolumn,prl,aps,amssymb]{revtex4}
\usepackage{epsf}
\begin{document}
\title{Hadamard type operations for qubits}
\author{Arpita Maitra and Preeti Parashar}
\affiliation{Physics and Applied Mathematics Unit,
Indian Statistical Institute,
203 B T Road, Kolkata 700 108, India,
Email: \{arpita\_r, parashar\}@isical.ac.in}
\begin{abstract}
We obtain the most general ensemble of qubits, for which it is
possible to design a universal Hadamard gate. These states when
geometrically represented on the Bloch sphere, give a new
trajectory. We further consider some Hadamard `type' of operations
and find ensembles of states for which such transformations hold.
Unequal superposition of a qubit and its orthogonal complement is
also investigated.

\end{abstract}
\maketitle
\newcommand{\qed}{\hfill \rule{2mm}{2mm}}
\newcommand{\pf}{{\bf Proof : }}
\newtheorem{definition}{Definition}
\newtheorem{algorithm}{Algorithm}
\newtheorem{construction}{Construction}
\newtheorem{theorem}{Theorem}
\newtheorem{question}{Question}
\newtheorem{lemma}{Lemma}
\newtheorem{proposition}{Proposition}
\newtheorem{remark}{Remark}
\newtheorem{corollary}{Corollary}
\newtheorem{example}{Example}
\newcommand{\binom}[2] {\mbox{$\left( { #1 \atop #2 } \right)$}}



\section{I. Introduction}
Qubits and quantum gates are the two basic building blocks of quantum
computers which are believed to be computationally stronger than their
classical counterparts.
One such important gate is the Hadamard gate which has found
wide applications in computer and communication
science  ~\cite{qNC02}.
There are a number of seminal papers in quantum computation and
information theory where Hadamard transform has
been used~\cite{qDJ92,BV93,BV97,qGR96, DD89}.
Shor's fast algorithm for factoring and discrete logarithm~\cite{Sh94}
are based on Fourier transform which is a generalization of the
Hadamard transform in higher dimensions.
Furthermore, the Toffoli and Hadamard gates comprise the simplest quantum
universal set of gates \cite{Shi02, DA03}. So, in order to achieve the full
power of quantum computation, one needs to add only the Hadamard
gate to the classical set. Thus, the role played by the Hadamard gate in
quantum algorithms is indeed significant.

Of late, Pati~\cite{PT02} has shown that one can not design
a universal Hadamard gate for an arbitrary unknown qubit.
Linearity, which is at the heart of quantum mechanics, does not allow
linear superposition of an unknown state $|\psi\rangle$ with its
orthogonal complement $|\psi_{\perp}\rangle$.
However, if one considers qubit states from the polar or equatorial great
circles on a Bloch sphere, then it is possible to design Hadamard type
of gates. By a Hadamard `type' gate we mean a unitary matrix
that is not exactly a Hadamard matrix. However, it still
creates an equal superposition (up to a sign or a phase) of a
qubit and its complement to produce two orthogonal states.
Very recently, Song et. al.~\cite{Son04}
have tried to implement the Hadamard gate in a probabilistic manner
for any unknown state chosen from a set
of linearly independent states.

Motivated by Pati's work, our primary aim in this paper is to
construct the most general class of qubit states, for which the
Hadamard gate can be designed in a deterministic way. This is
achieved in Sec. II, by imposing restrictions ( due to linearity )
on a completely arbitrary unknown quantum state. States from this
set are geometrically represented on the three - dimensional unit
sphere known as the Bloch sphere. In Sec. III, we show that certain
Hadamard `type' transformations are indeed possible for arbitrary
states when partial information is available. A Hadamard type gate
is obtained for qubits chosen from, not only the polar great circle
but also from any polar circle. We also demonstrate with an example,
that there is a unique class of states (up to isomorphism)
associated with a particular gate, satisfying a fixed
transformation. As for the second Hadamard type of transformation,
which is related to the states lying on the equatorial great circle,
a new ensemble of states is found. In Sec. IV, unequal superposition
of a qubit with its orthogonal complement is investigated. This is a
generalization of the usual Hadamard transformation when the two
amplitudes are not equal. In this context, many new classes of
quantum states are found for which the unequal superposition
works.Summary and Concluding remarks are made in Sec. V.

\section{II. Hadamard Transform for Special Qubits}

The Hadamard transform $H$, which is a one qubit gate,
rotates the two computational basis vectors $|0\rangle$ and
$|1\rangle$
to two other orthogonal vectors $\frac{1}{\sqrt{2}} (|0\rangle + |1\rangle)$
and $\frac{1}{\sqrt{2}} (|0\rangle - |1\rangle)$, respectively. Thus
it creates an equal superposition of the amplitudes of the state and its
orthogonal. The matrix representation of the Hadamard gate is given by
$H = \frac{1}{\sqrt{2}}
\left[\begin{array}{cr}
1 & 1 \\
1 & -1 \\
\end{array}\right]$.

The question we ask in this paper is: What is the most general set of
qubit states $\{|\psi\rangle, |\psi_{\perp}\rangle\}$, such that the
application of
the Hadamard gate $H$ takes them to two other orthogonal states
$\frac{1}{\sqrt{2}} (|\psi\rangle + |\psi_{\perp}\rangle)$
and $\frac{1}{\sqrt{2}} (|\psi\rangle - |\psi_{\perp}\rangle)$
respectively? If $|\psi\rangle$ is completely arbitrary and
unknown, then such a universal
Hadamard gate does not exist \cite{PT02}. So,
we shall obtain a special class of qubit states such that
\begin{equation}
\label{eq1}
H(|\psi\rangle) = \frac{1}{\sqrt{2}} (|\psi\rangle + |\psi_{\perp}\rangle),
H(|\psi_{\perp}\rangle) = \frac{1}{\sqrt{2}}
(|\psi\rangle - |\psi_{\perp}\rangle).
\end{equation}

\begin{figure*}
\centerline{\epsfxsize=5in\epsfbox{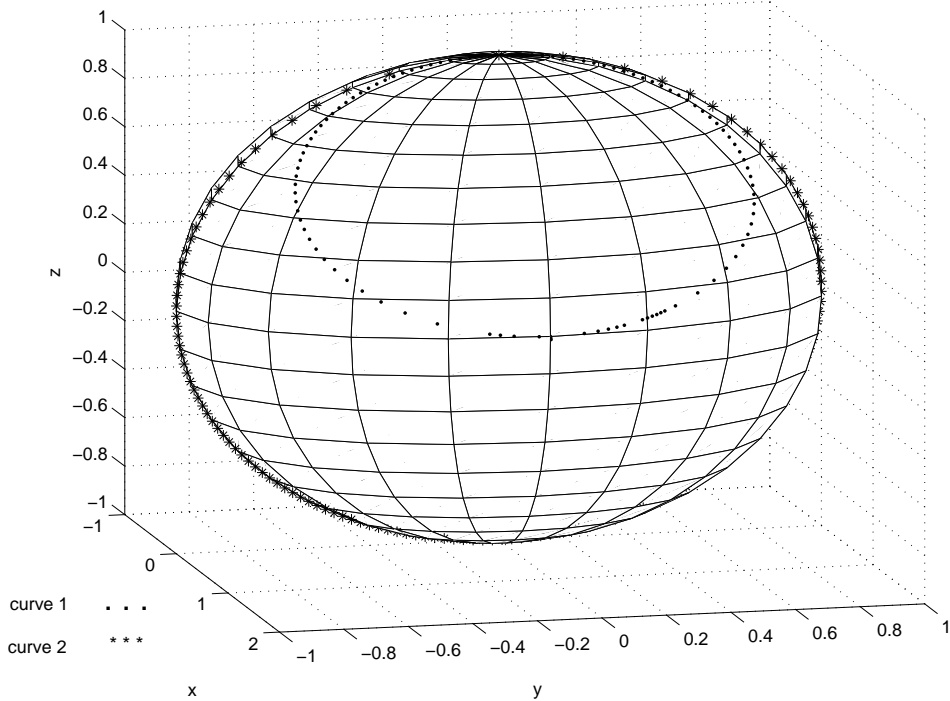}}
\caption{Points on Bloch sphere in reference to Theorem~\ref{th1} and
Theorem~\ref{th3}}
\label{fig1}
\end{figure*}

We start by considering a completely arbitrary, unknown qubit state
$|\psi\rangle = a|0\rangle + b |1\rangle$ and its orthogonal
complement $|\psi_{\perp}\rangle = b^* |0\rangle - a^* |1\rangle$.
Here $a, b$ are complex numbers obeying the normalization condition
$|a|^2 + |b|^2 = 1$. The operator $H$ acts linearly, i.e.,
$H(|\psi\rangle) = a H(|0\rangle) + b H(|1\rangle)$. In what
follows, we show that this restricts the form of $|\psi\rangle$ to
$(\alpha + i\beta) |0\rangle + \alpha |1\rangle$, with $2\alpha^2 +
\beta^2 = 1$, where $\alpha$ and $\beta$ are real.

Now we substantiate the assertions made herein above. Ideally, from
the Hadamard transformation we obtain
\begin{eqnarray}
H(|\psi\rangle) &=& \frac{1}{\sqrt{2}} (|\psi\rangle + |\psi_{\perp}\rangle)
\nonumber \\
\ &=& \frac{1}{\sqrt{2}}
(a |0\rangle + b |1\rangle + b^* |0\rangle - a^* |1\rangle) \nonumber \\
\ &=& \frac{(a+b^*) |0\rangle - (a^*-b) |1\rangle}{\sqrt{2}}, \nonumber
\end{eqnarray}
and from linearity we get
\begin{eqnarray}
H(|\psi\rangle) &=& a H(|0\rangle) + b H(|1\rangle) \nonumber \\
\ &=& a \frac{|0\rangle + |1\rangle}{\sqrt{2}} +
      b \frac{|0\rangle - |1\rangle}{\sqrt{2}} \nonumber \\
\ &=& \frac{(a+b) |0\rangle + (a-b) |1\rangle}{\sqrt{2}}. \nonumber
\end{eqnarray}
These two expressions should be equal.
Hence $a + b^* = a + b$, i.e., $b = b^*$. Thus $b$ is real. Let $b = \alpha$,
where $\alpha$ is a real number.
Moreover, $-(a^* - b) = (a - b)$, i.e.,
$a+a^* = 2b$. So the real part of $a$ is $\alpha$. Let $a = \alpha + i\beta$,
where $\beta$ is a real number.
Thus $|\psi\rangle$ is of the form $(\alpha + i\beta) |0\rangle +
\alpha |1\rangle$.
Clearly $-\frac{1}{\sqrt{2}} \leq \alpha \leq \frac{1}{\sqrt{2}}$, since
$\beta^2 = 1 - 2 \alpha^2$.
Therefore, $|\psi\rangle$ has complex as well as real amplitudes when expressed
in the computational basis $\{|0\rangle, |1\rangle \}$; the real
parts of which are equal.

It can be easily checked, in a similar fashion, that if we consider
$H(|\psi_{\perp}\rangle)$, we get $b = b^*$ (i.e., $b$ is real) and
$a+a^* = 2b^* = 2b$ (as $b$ is real) leading to the same result.
Thus $|\psi_{\perp}\rangle = b^* |0\rangle - a^* |1\rangle$ is restricted
to the form $\alpha |0\rangle - (\alpha - i\beta) |1\rangle$.

Our next task is to map the states from this ensemble to points on
the Bloch sphere. But before attempting to do this, we give a brief
pedagogical description of how to geometrically represent a general
qubit state on the Bloch sphere. Consider $|\psi\rangle = a
|0\rangle + b |1\rangle$. Since $a$ and $b$ are complex,  assume $a
= r_1 e^{i\gamma}, b = r_2 e^{i(\gamma+\phi)}$. Then $|a| = r_1, |b|
= r_2$. Let $r_1 = \cos{\frac{\theta}{2}}, r_2 =
\sin{\frac{\theta}{2}}$. Hence, $a = \cos{\frac{\theta}{2}}
e^{i\gamma}, b = \sin{\frac{\theta}{2}} e^{i(\gamma+\phi)}$, where
$\theta, \phi$ and $\gamma$ are real.

Thus, any qubit $|\psi\rangle = a |0\rangle + b |1\rangle$ can be
written as $e^{i\gamma}(\cos{\frac{\theta}{2}} |0\rangle + e^{i\phi}
\sin{\frac{\theta}{2}} |1\rangle)$. Further, two qubits
$e^{i\gamma}(\cos{\frac{\theta}{2}} |0\rangle + e^{i\phi}
\sin{\frac{\theta}{2}} |1\rangle)$ and $(\cos{\frac{\theta}{2}}
|0\rangle + e^{i\phi} \sin{\frac{\theta}{2}} |1\rangle)$ are treated
on equal footing under measurement, since they differ only by an
overall phase factor which has no observable effect.

The qubit $(\cos{\frac{\theta}{2}} |0\rangle +
e^{i\phi} \sin{\frac{\theta}{2}} |1\rangle)$ is mapped to a point
$(1, \theta, \phi)$ on the unit Bloch sphere. Here $\theta$ and $\phi$
are the usual polar and azimuthal angles respectively, and they are
related to the cartesian coordinates $(x, y, z)$ through the usual relations
$x = \cos{\phi} \sin{\theta}$,
$y = \sin{\phi} \sin{\theta}$, $z = \cos{\theta}$.
If we fix $\phi = 0$, then we obtain states of the form
$\cos{\frac{\theta}{2}} |0\rangle + \sin{\frac{\theta}{2}} |1\rangle$ and
$\cos{\frac{\theta}{2}} |1\rangle - \sin{\frac{\theta}{2}} |0\rangle$
for $0 \leq \theta \leq \pi$. These lie on the polar great circle of
the Bloch sphere.
On the other hand, for $\theta = \pi /2$, one obtains states of the form
$\frac{1}{\sqrt{2}}(|0\rangle + e^{i\phi} |1\rangle)$ and
$\frac{1}{\sqrt{2}}(|1\rangle - e^{-i\phi} |0\rangle)$
for $0 \leq \phi \leq 2\pi$, which lie on the equatorial great circle.

Now we can conveniently plot the states from our special ensemble
$|\psi\rangle = (\alpha + i\beta) |0\rangle + \alpha |1\rangle$.
Bringing it to the desired form, it is clear that
$e^{i\gamma} = \frac{\alpha + i\beta}{\sqrt{\alpha^2 + \beta^2}},
\cos{\frac{\theta}{2}} = \sqrt{\alpha^2 + \beta^2},
\sin{\frac{\theta}{2}} = \alpha,
e^{i\phi} = \frac{\sqrt{\alpha^2 + \beta^2}}{\alpha + i\beta}$.
We thus arrive at the following identification:
\begin{eqnarray}
\label{al1}
x &=& \cos{\phi}\sin{\theta} = 2\alpha^2, \nonumber \\
y &=& \sin{\phi} \sin{\theta} = - 2\alpha \sqrt{1-2\alpha^2}, \nonumber \\
z &=& \cos{\theta} = 1 - 2\alpha^2.
\end{eqnarray}

These points are represented by curve 1 on the Bloch sphere.

Next, we consider
$|\psi\rangle = a|0\rangle + b |1\rangle$ and the other orthogonal
complement
$|\psi_{\perp}\rangle = - b^* |0\rangle + a^* |1\rangle$, which differs
from the first one just by an overall negative sign.
This yields that
$|\psi\rangle$ must be of the form
$(\alpha + i\beta) |0\rangle + i\beta |1\rangle$,
with $\alpha^2 = 1 - 2\beta^2$
where $-\frac{1}{\sqrt{2}} \leq \beta \leq \frac{1}{\sqrt{2}}$, since
$\alpha$ is real.
Therefore, in the computational basis, the qubit state $|\psi\rangle$ has
complex and imaginary
amplitudes; the imaginary parts of which are equal.
As for $|\psi_{\perp}\rangle = -b^* |0\rangle + a^* |1\rangle$,
it assumes the form $i\beta |0\rangle + (\alpha - i\beta) |1\rangle$.

For the qubits of the form $(\alpha + i\beta) |0\rangle + i\beta |1\rangle$,
we have $e^{i\gamma} = \frac{\alpha + i\beta}{\sqrt{\alpha^2 + \beta^2}},
\cos{\frac{\theta}{2}} = \sqrt{\alpha^2 + \beta^2},
\sin{\frac{\theta}{2}} = \beta,
e^{i\phi} = i\frac{\sqrt{\alpha^2 + \beta^2}}{\alpha + i\beta}$.
Thus on the Bloch sphere:
\begin{eqnarray}
\label{al2}
x &=& \cos{\phi}\sin{\theta} = 2\beta^2, \nonumber \\
y &=& \sin{\phi} \sin{\theta} = 2\beta \sqrt{1-2\beta^2}, \nonumber \\
z &=& \cos{\theta} = 1 - 2\beta^2.
\end{eqnarray}

It is immediately clear that a point represented by Eq(\ref{al1}), for a
particular value of $\alpha$, is exactly
equal to the one obtained from Eq(\ref{al2}) for the same value of
$(- \beta)$. This implies that these two ensembles
give the same trajectory on the Bloch sphere. Hence, we shall
consider them to be isomorphic to each other.
Our result is thus summarized in the following theorem.
\begin{theorem}
\label{th1}
The most general qubit states for which it is possible to design a
universal Hadamard gate satisfying
Eq(\ref{eq1}) are given by
$\{{|\psi\rangle, |\psi_{\perp}\rangle} || |\psi\rangle = (\alpha +
i\beta)
|0\rangle + \alpha |1\rangle ;
|\psi_{\perp}\rangle = \alpha |0\rangle - (\alpha - i\beta) |1\rangle\}$
where $\alpha, \beta$ are real such that
$2\alpha^2 + \beta^2 = 1$ and
$-\frac{1}{\sqrt{2}} \leq \alpha \leq \frac{1}{\sqrt{2}}$.
\end{theorem}

Note that if we choose $\alpha = 0$, then from Eq(\ref{al1}), we obtain
the point $(0, 0, 1)$ on the Bloch sphere (i.e., north pole), which can be
identified with
the computational basis state $|0\rangle$
on curve 1.

In a similar fashion, the trajectory of $|\psi_\perp\rangle$
can be sketched, which would lie on the other side of
the Bloch sphere (not visible in the figure). It can be checked that
the orthogonal state $|1\rangle$ would be one of its points
$(0, 0, -1)$ (i.e., south pole).

We demonstrate that this trajectory has some intersection points with the
equatorial great circle also. To this end, for $\theta = \frac{\pi}{2}$,
$z = \cos{\theta} = 0 = 1 - 2\alpha^2$, i.e., $\alpha = \pm \frac{1}{\sqrt{2}}$,
and $2\alpha^2 + \beta^2 = 1$, i.e., $\beta = 0$. Substituting these values
we get $|\psi\rangle = (\alpha + i \beta) |0\rangle + \alpha |1\rangle
= \pm \frac{1}{\sqrt{2}}(|0\rangle + |1\rangle)$ and
$|\psi_\perp\rangle = \alpha |0\rangle + (\alpha - i\beta) |1\rangle
= \pm \frac{1}{\sqrt{2}}(|0\rangle - |1\rangle)$. For the second case,
$\beta = \pm \frac{1}{\sqrt{2}}$ and $\alpha = 0$. This yields
$|\psi\rangle =
= \pm \frac{i}{\sqrt{2}}(|0\rangle + |1\rangle)$ and
$|\psi_\perp\rangle = = \pm \frac{i}{\sqrt{2}}(|0\rangle - |1\rangle)$.

Therefore,if one chooses any qubit $|\psi\rangle$ from curve 1 in
the figure, and takes its orthogonal complement, then the Hadamard
transformation works perfectly well to generate the superposition.
To see it explicitly, take $|\psi\rangle = (\alpha + i\beta)
|0\rangle + \alpha |1\rangle$ and $|\psi_\perp\rangle = \alpha
|0\rangle - (\alpha - i\beta) |1\rangle$. $H$ rotates $|\psi\rangle$
to
\begin{eqnarray}
H|\psi\rangle &=&
\frac{1}{\sqrt{2}}
\left[\begin{array}{cr}
1 & 1 \\
1 & -1 \\
\end{array}\right]
\left[\begin{array}{c}
\alpha + i\beta \\
\alpha \\
\end{array}\right]
= \frac{1}{\sqrt{2}}
\left[\begin{array}{c}
\alpha + i\beta + \alpha\\
\alpha + i\beta - \alpha\\
\end{array}\right] \nonumber \\
\ &=& \frac{1}{\sqrt{2}}
\left[\begin{array}{c}
(\alpha + i\beta) + (\alpha)\\
(\alpha) - (\alpha - i\beta)\\
\end{array}\right]
= \frac{1}{\sqrt{2}} (|\psi\rangle + |\psi_{\perp}\rangle), \nonumber
\end{eqnarray}

while it acts on $|\psi_{\perp}\rangle$ to give
\begin{eqnarray}
H|\psi_{\perp}\rangle &=&
\frac{1}{\sqrt{2}}
\left[\begin{array}{cr}
1 & 1 \\
1 & -1 \\
\end{array}\right]
\left[\begin{array}{c}
\alpha \\
-(\alpha - i\beta) \\
\end{array}\right] \nonumber \\
\ &=& \frac{1}{\sqrt{2}}
\left[\begin{array}{c}
\alpha - (\alpha - i\beta)\\
\alpha + (\alpha - i\beta)\\
\end{array}\right] \nonumber \\
\ &=& \frac{1}{\sqrt{2}}
\left[\begin{array}{c}
(\alpha + i\beta) - (\alpha)\\
(\alpha) + (\alpha - i\beta)\\
\end{array}\right]
= \frac{1}{\sqrt{2}} (|\psi\rangle - |\psi_{\perp}\rangle).\nonumber
\end{eqnarray}

Alternatively, one can also prove the above theorem by unitarity,
as was done in~\cite{PT02} for the general case.
Take any two quantum states $\{|\psi^{(k)}\rangle, |\psi^{(l)}\rangle\}$
from this special ensemble, and their complement states
$\{|\psi_{\perp}^{(k)}\rangle, |\psi_{\perp}^{(l)}\rangle\}$.
Applying the Hadamard transformation (\ref{eq1}) on them and taking
inner product, we get

\begin{eqnarray}
\label{eqip1}
\langle\psi^{(k)}
|\psi^{(l)}\rangle &=& \frac{1}{2}(\langle\psi^{(k)}
|\psi^{(l)}\rangle + \langle\psi^{(k)}
|\psi_{\perp}^{(l)}\rangle + \langle\psi_{\perp}^{(k)}
|\psi^{(l)}\rangle \nonumber \\
\ &+& \langle\psi_{\perp}^{(k)}|\psi_{\perp}^{(l)}\rangle)
\end{eqnarray}

\begin{eqnarray}
\label{eqip2}
\langle\psi_{\perp}^{(k)}
|\psi_{\perp}^{(l)}\rangle &=& \frac{1}{2}(\langle\psi^{(k)}
|\psi^{(l)}\rangle - \langle\psi^{(k)}
|\psi_{\perp}^{(l)}\rangle - \langle\psi_{\perp}^{(k)}
|\psi^{(l)}\rangle \nonumber \\
\ &+& \langle\psi_{\perp}^{(k)}|\psi_{\perp}^{(l)}\rangle).
\end{eqnarray}

Any two qubits from this ensemble obey the conjugation rules
\begin{eqnarray}
\label{eqip3}
\langle\psi^{(k)}| \psi_{\perp}^{(l)}\rangle &=&
- \langle\psi_{\perp}^{(k)}| \psi^{(l)}{\rangle}^*
=  \langle\psi_{\perp}^{(k)}| \psi^{(l)}\rangle, \nonumber \\
\langle\psi^{(k)}| \psi^{(l)}\rangle &=&
\langle\psi_{\perp}^{(k)}| \psi_{\perp}^{(l)}{\rangle}^*.
\end{eqnarray}

Substituting these conditions in the above inner product relations,
it is straightforward to check that the inner
product is preserved. Hence, a universal Hadamard gate exists for any
qubit chosen from this special class.

\section{III. Hadamard Type Transforms}
In this section, we consider some operations which are not exactly
Hadamard transforms, but similar, in the sense that they produce
equal superposition of the amplitudes up to a sign or a phase.
These have been discussed by Pati, in the context of qubits from
the polar and equatorial great circles. Here, we elaborate more on these
transformations and present some general results.

\subsection{Polar Type Transformation}

Any two orthogonal vectors on the polar great circle,
$|\psi\rangle =
\cos{\frac{\theta}{2}} |0\rangle + \sin{\frac{\theta}{2}} |1\rangle$ and
$|\psi_{\perp}\rangle =
\cos{\frac{\theta}{2}} |1\rangle - \sin{\frac{\theta}{2}} |0\rangle$,
can be shown to transform as \cite{PT02}
\begin{equation}
\label{eq1a}
U(|\psi\rangle) = \frac{1}{\sqrt{2}} (|\psi\rangle +
|\psi_{\perp}\rangle),
U(|\psi_{\perp}\rangle) = \frac{1}{\sqrt{2}} (|\psi_{\perp}\rangle -
|\psi\rangle).
\end{equation}
This differs from the usual Hadamard transformation by an overall sign
in the second part. The appropriate unitary operator $U$ which does the job is
denoted by

$H_P = \sigma_x H =
\frac{1}{\sqrt{2}}
\left[\begin{array}{cr}
1 & -1 \\
1 & 1 \\
\end{array}\right]$, where $\sigma_x$ is the Pauli flip matrix
$\left[\begin{array}{cr}
0 & 1 \\
1 & 0 \\
\end{array}\right]$.

We now extend this result to vectors from any polar circle.
Take $|\psi\rangle = \cos{\frac{\theta}{2}} |0\rangle +
e^{i\phi} \sin{\frac{\theta}{2}} |1\rangle$ on any of the polar
circles, and its orthogonal complement
$|\psi_{\perp}\rangle = -\sin{\frac{\theta}{2}} |0\rangle
+ e^{i\phi} \cos{\frac{\theta}{2}} |1\rangle$. Then, for any $\phi$, we
can construct a unitary operator

$H_G^{\phi} =
\frac{1}{\sqrt{2}}
\left[\begin{array}{cr}
1 & -e^{-i\phi} \\
e^{i\phi} & 1 \\
\end{array}\right]$,
such that Eq(\ref{eq1a}) is satisfied, i.e.,
$H_G^{\phi} |\psi\rangle = \frac{1}{\sqrt{2}}
(|\psi\rangle + |\psi_{\perp}\rangle)$ and
$H_G^{\phi} |\psi_\perp\rangle = \frac{1}{\sqrt{2}}
(|\psi_\perp\rangle - |\psi\rangle)$. For $\phi = 0$,
$H_G^{\phi}$ reduces to $H_P$, thereby covering the polar great circle
case.

\begin{theorem}
\label{th2}
For any state $|\psi\rangle = \cos{\frac{\theta}{2}} |0\rangle +
e^{i\phi} \sin{\frac{\theta}{2}} |1\rangle$ and its orthogonal state
$|\psi_{\perp}\rangle = -\sin{\frac{\theta}{2}} |0\rangle
+ e^{i\phi} \cos{\frac{\theta}{2}} |1\rangle$, it is
possible to design a Hadamard type gate $H_G^{\phi}$ that satisfies
the transformation (\ref{eq1a}), once $\phi$ is known.
\end{theorem}

It is clear that the unitary operator $U$,
satisfying Eq(\ref{eq1a}), depends on the type of states chosen.
For instance, we get two different gates above, depending on
whether $\{|\psi\rangle, |\psi_{\perp}\rangle\}$ belongs to the polar
great circle or some other polar circle.
Therefore, if one fixes the operator $U$, then one can show
that there is a unique ensemble of states satisfying the transformation
(\ref{eq1a}). For this purpose, let us consider the gate $H_P$, and find
the associated class of states for which it works.

We follow our previous procedure of taking
$|\psi\rangle = a|0\rangle + b |1\rangle$ and its orthogonal
complement $|\psi_{\perp}\rangle = b^* |0\rangle - a^* |1\rangle$.
Linearity yields,
\begin{eqnarray}
H_P(|\psi\rangle) &=& aH_P(|0\rangle) + bH_P(|1\rangle)\\
\ &=&
\frac{a}{\sqrt{2}} (|0\rangle + |1\rangle) +
\frac{b}{\sqrt{2}} (|1\rangle - |0\rangle) \nonumber \\
\ &=& \frac{(a-b) |0\rangle - (a+b) |1\rangle}{\sqrt{2}}. \nonumber
\end{eqnarray}
On the other hand,
\begin{eqnarray}
H(|\psi\rangle) &=& \frac{1}{\sqrt{2}}(
|\psi\rangle+|\psi_\perp\rangle) \nonumber \\
\ &=& \frac{a|0\rangle + b|1\rangle}{\sqrt{2}} +
      \frac{b^*|0\rangle - a^*|1\rangle}{\sqrt{2}} \nonumber \\
\ &=& \frac{(a+b^*) |0\rangle + (b-a^*) |1\rangle}{\sqrt{2}} \nonumber
\end{eqnarray}
Thus, $a - b = a + b^*$, i.e., $b+b^* = 0$. So $b$ is imaginary.
Moreover, $a + b = b - a^*$, i.e., $a+a^* = 0$. Therefore, $a$ is also
imaginary.
Hence, given $\alpha, \beta$ real, we get
$|\psi\rangle = i\alpha |0\rangle + i \beta |1\rangle$ and
$|\psi_{\perp}\rangle = -i\beta |0\rangle + i \alpha |1\rangle$.
Rewriting $|\psi\rangle = i(\alpha |0\rangle + \beta |1\rangle)$
and representing it on the Bloch sphere, one can readily check that
$e^{i\gamma} = i$, $\cos{\frac{\theta}{2}} = \alpha$,
$\sin{\frac{\theta}{2}} = \beta$, $e^{i\phi} = 1$. The resulting trajectory
is that of the polar great circle.

However, if we had taken
$|\psi_{\perp}\rangle = -b^* |0\rangle + a^* |1\rangle$,
we would have got $|\psi\rangle$ of the form $\alpha |0\rangle +
\beta |1\rangle$, which again are the states on the polar great circle.
We thus conclude that, up to isomorphism, this is the only class of qubit
states which
transforms according to Eq(\ref{eq1a}), under the action of the gate
$H_P$. Alternatively, one can also fix the states and determine the
corresponding gate uniquely.
\subsection{Equatorial Type Transformation}

The second kind of operation discussed in~\cite{PT02} is that of an equal
superposition of amplitudes up to a phase such that
\begin{equation}
\label{eq2}
U(|\psi\rangle) = \frac{1}{\sqrt{2}} (|\psi\rangle + i|\psi_{\perp}\rangle),
U(|\psi_{\perp}\rangle) = \frac{1}{\sqrt{2}}
(i|\psi\rangle + |\psi_{\perp}\rangle).
\end{equation}

This alternative universal definition of a Hadamard type gate, has the
advantage that it is invariant under the
interchange of $|\psi\rangle$ and $|\psi_{\perp}\rangle$.
Vectors of the form
$|\psi(\phi)\rangle = H(\cos\frac{\phi}{2} |0\rangle) -
i \sin\frac{\phi}{2} |1\rangle) =
\frac{1}{\sqrt{2}} e^{-i\phi /2} (|0\rangle + e^{i\phi} |1\rangle)$
and the corresponding orthogonal
$|\psi_{\perp}(\phi)\rangle = H(i\sin\frac{\phi}{2} |0\rangle) -
\cos\frac{\phi}{2} |1\rangle) =
\frac{1}{\sqrt{2}} e^{i\phi /2} (|1\rangle - e^{-i\phi} |0\rangle)$
chosen from the equatorial great circle
satisfy this transformation
provided the unitary matrix is
$H_E =
\frac{1}{\sqrt{2}}
\left[\begin{array}{cr}
1-i & 0 \\
0 & 1+i \\
\end{array}\right]$.
We wish to clarify here that the matrix $H_E$ presented in \cite{PT02}
does not work for the states considered, and the correct form of $H_E$
should essentially be what we have given above.

Our next task is to find the most general class of states satisfying
the phase
dependent transformation (\ref{eq2}), provided the computational basis
vectors $\{|0\rangle, |1\rangle\}$ also transform in the same fashion,
i.e., to
$\frac{1}{\sqrt{2}}(|0\rangle + i |1\rangle)$ and
$\frac{1}{\sqrt{2}}(i |0\rangle + |1\rangle)$, respectively.

Thus fixing the unitary operator as
$U = \frac{1}{\sqrt{2}}
\left[\begin{array}{cr}
1 & i \\
i & 1 \\
\end{array}\right]$,
we obtain conditions on the form of $|\psi\rangle$ and
$|\psi_{\perp}\rangle$.
Following the earlier procedure, we
assume that $|\psi\rangle = a|0\rangle + b |1\rangle$ and
$|\psi_{\perp}\rangle = b^* |0\rangle - a^* |1\rangle$. Then using
linearity
of the operation, we find that $|\psi\rangle$ must be of the form
$i\alpha |0\rangle + \beta |1\rangle$. The complement
$|\psi_{\perp}\rangle$ is restricted to
$\beta |0\rangle + i\alpha |1\rangle$. Now $|\psi\rangle$ can be written as
$i\sqrt{\alpha^2+\beta^2} (\frac{\alpha}{\sqrt{\alpha^2+\beta^2}} |0\rangle
- i\frac{\beta}{\sqrt{\alpha^2+\beta^2}} |1\rangle)$, i.e.,
$e^{i\gamma}(\cos{\frac{\theta}{2}} |0\rangle +
e^{i\frac{3\pi}{2}} \sin{\frac{\theta}{2}} |1\rangle)$.
On the Bloch sphere,
$x = \cos{\phi}\sin{\theta}
= 0, y = \sin{\phi} \sin{\theta} = 2\alpha \sqrt{1-\alpha^2},
z = \cos{\theta} = 2\alpha^2 - 1$, where
$-1 \leq \alpha \leq 1$.

Similarly considering the second complement $|\psi_{\perp}\rangle =
-b^* |0\rangle + a^* |1\rangle$, and using linearity of the
operation, we get $|\psi\rangle = \alpha |0\rangle + i\beta
|1\rangle$ and $|\psi_{\perp}\rangle = i\beta |0\rangle + \alpha
|1\rangle$. Identifying with the Bloch sphere picture,
$|\psi\rangle$ can be written as $\sqrt{\alpha^2+\beta^2}
(\frac{\alpha}{\sqrt{\alpha^2+\beta^2}} |0\rangle +
i\frac{\beta}{\sqrt{\alpha^2+\beta^2}} |1\rangle)$, i.e.,
$e^{i\gamma}(\cos{\frac{\theta}{2}} |0\rangle + e^{i\frac{\pi}{2}}
\sin{\frac{\theta}{2}} |1\rangle)$. Therefore, on the Bloch sphere,
$x = \cos{\phi}\sin{\theta} = 0, y = \sin{\phi} \sin{\theta} =
2\alpha \sqrt{1-\alpha^2}, z = \cos{\theta} = 2\alpha^2 - 1$, where
$-1 \leq \alpha \leq 1$.

Hence $|\psi\rangle$, when expressed in computational basis,
is made up of one real and one imaginary amplitude.
As expected, the above two ensembles give the same trajectory,
represented by curve 2 in the figure.
We thus have the following result.
\begin{theorem}
\label{th3}
It is possible to design a universal Hadamard type gate $U$,
satisfying the transformation (\ref{eq2}),
for any state of the form
$|\psi\rangle =
i\alpha |0\rangle + \beta |1\rangle$ and its orthogonal complement
$|\psi_{\perp}\rangle =
\beta |0\rangle + i\alpha |1\rangle$, where $\alpha, \beta$ are real
such that $\alpha^2 + \beta^2 = 1$ and $-1 \leq \alpha \leq 1$.
\end{theorem}

One can check explicitly that
$U(|\psi\rangle) =
\frac{1}{\sqrt{2}}
\left[\begin{array}{cr}
1 & i \\
i & 1 \\
\end{array}\right]$
$\left[\begin{array}{c}
i\alpha \\
\beta \\
\end{array}\right]$
$= \frac{1}{\sqrt{2}}
\left[\begin{array}{c}
i(\alpha + \beta)\\
-\alpha + \beta\\
\end{array}\right]
= \frac{1}{\sqrt{2}} (|\psi\rangle + i|\psi_{\perp}\rangle)$.
Similarly,

$U(|\psi_{\perp}\rangle) =
\frac{1}{\sqrt{2}}
\left[\begin{array}{cr}
1 & i \\
i & 1 \\
\end{array}\right]$
$\left[\begin{array}{c}
\beta \\
i\alpha \\
\end{array}\right]$
$= \frac{1}{\sqrt{2}}
\left[\begin{array}{c}
\beta - \alpha\\
i(\beta + \alpha)\\
\end{array}\right]
= \frac{1}{\sqrt{2}} (i|\psi\rangle + |\psi_{\perp}\rangle)$.
Interestingly, this trajectory cuts the equatorial great circle when
$z = 0$, i.e., $\alpha = \pm \frac{1}{\sqrt{2}}$. These intersection
points imply that there
are quantum states (and their orthogonals ) from this
ensemble which also
belong to the equatorial great circle. Let us find out these states.

For $\theta = \frac{\pi}{2}$, $\alpha = \pm \frac{1}{\sqrt{2}}$, and from
normalization condition, $\beta = \pm \frac{1}{\sqrt{2}}$. Substituting these
values in $|\psi\rangle$ and $|\psi_\perp\rangle$ of Theorem~\ref{th3}, we get
$|\psi\rangle = \pm \frac{1}{\sqrt{2}} (i |0\rangle \pm |1\rangle)$ and
$|\psi_\perp\rangle = \pm \frac{1}{\sqrt{2}} (|0\rangle \pm i|1\rangle)$.
One can similarly find the states corresponding to the second orthogonal
complement.

\section{IV. Unequal Superposition}
We shall now focus our attention on unequal superposition of the
amplitudes of a qubit state. Like the equal superposition case, it
is impossible to create unequal superposition of an arbitrary
unknown qubit with its complement state \cite{PT02}. Our task
therefore, is to obtain special classes of states for which such a
superposition would be possible. This can be regarded as a
generalized version of the usual Hadamard transformation and is
given by
\begin{equation}
\label{equ1}
U(|\psi\rangle) = p|\psi\rangle + q|\psi_{\perp}\rangle,
U(|\psi_{\perp}\rangle) = q^*|\psi\rangle - p^*|\psi_{\perp}\rangle.
\end{equation}
Here, $p, q$ are known complex numbers with $|p|^2 + |q|^2 = 1$.
We again demand that
$|\psi\rangle$ and $|\psi_{\perp}\rangle$ transform like
the computational basis vectors $|0\rangle$ and
$|1\rangle$ respectively.
Thus $U$ can be fixed as
$\left[\begin{array}{cr}
p & q^* \\
q & -p^* \\
\end{array}\right]$. Repeating the linearity
procedure,
take $|\psi\rangle = a|0\rangle + b |1\rangle$ and
$|\psi_{\perp}\rangle = b^* |0\rangle - a^* |1\rangle$.
Ideally we should have
\begin{eqnarray}
U(|\psi\rangle) &=& p|\psi\rangle + q|\psi_{\perp}\rangle
\nonumber \\
\ &=& p(a |0\rangle + b |1\rangle) +
    q(b^* |0\rangle - a^* |1\rangle) \nonumber \\
\ &=& (pa+qb^*) |0\rangle + (pb - qa^*) |1\rangle. \nonumber
\end{eqnarray}
On the other hand, from linearity we get
\begin{eqnarray}
U(|\psi\rangle) &=& a U(|0\rangle) + b U(|1\rangle) \nonumber \\
\ &=& a (p|0\rangle + q|1\rangle) +
      b (q^*|0\rangle - p^*|1\rangle) \nonumber \\
\ &=& (ap+bq^*) |0\rangle + (aq-bp^*) |1\rangle \nonumber
\end{eqnarray}
Hence $pa + qb^* = pa + q^*b$, i.e., $qb^* = q^*b = (qb^*)^*$.
Thus $qb^*$ is real, which implies that
\begin{enumerate}
\item both $q, b$ are real, or
\item both $q, b$ are imaginary, or
\item both $q, b$ are complex, with the constraint
$\frac{q_1}{q_2} = \frac{b_1}{b_2}$.
(Here, any complex number $z = (a, b, q, p)$ has been written as
$z= z_1 + i z_2$).
\end{enumerate}
Further, $pb - qa^* = qa - p^*b$, i.e., $q(a + a^*) = b(p + p^*)$,
so, $q \cdot Re(a) = b \cdot Re(p)$, i.e., $Re(a) = \frac{b}{q} \cdot
Re(p)$.

Therefore, $|\psi\rangle$ and $|\psi_{\perp}\rangle$
are restricted to the form
$|\psi\rangle = (\frac{b}{q} \cdot Re(p) + ia_2) |0\rangle + b
|1\rangle$ and
$|\psi_{\perp}\rangle = b^* |0\rangle -
(\frac{b}{q} Re(p) - i a_2) |1\rangle$.
Depending on whether $q$ and $b$ are both real, or imaginary or
complex ( with $\frac{q_1}{q_2} = \frac{b_1}{b_2}$), we get different
classes of states for which the unequal superposition transformation
(\ref{equ1}) holds. For the special value of
$p = q = \frac{1}{\sqrt{2}}$, this ensemble goes over to the set
of states
$|\psi\rangle = (b + ia_2) |0\rangle + b |1\rangle$ (i.e., complex and
real amplitudes such that real parts are the same)
obtained in Sec. II. Also, the associated Hadamard matrix $H$, satisfying
the transformation (\ref{eq1}), can be recovered for these
values of $p,q$ from the above $U$.
Now, analogous to
the previous section, we concentrate below, on two specific unequal
superposition transformations.

\subsection{Unequal Polar Type Transformation}
According to the prescription outlined in~\cite{PT02}, for the vectors on
the polar great circle, one can find a unitary gate
$U_P =
\left[\begin{array}{cr}
p & -q \\
q & p \\
\end{array}\right]$, where $p^2 + q^2 = 1$ and $p, q$ are now real.
In this case $|\psi\rangle =
\cos{\frac{\theta}{2}} |0\rangle + \sin{\frac{\theta}{2}} |1\rangle$ and
$|\psi_{\perp}\rangle =
\cos{\frac{\theta}{2}} |1\rangle - \sin{\frac{\theta}{2}} |0\rangle$
and they transform as
\begin{equation}
\label{equ2}
U_P (|\psi\rangle) = q|\psi_{\perp}\rangle + p|\psi\rangle,
U_P (|\psi_{\perp}\rangle) = p|\psi_{\perp}\rangle - q|\psi\rangle.
\end{equation}
This is almost similar to Eq(\ref{equ1}), up to an overall
sign (when $p, q$ are real).

Now we present a generalization of this result. Take a qubit
$|\psi\rangle = \cos{\frac{\theta}{2}} |0\rangle +
e^{i\phi} \sin{\frac{\theta}{2}} |1\rangle$ on any polar
circle,
and the orthogonal qubit
$|\psi_{\perp}\rangle = -\sin{\frac{\theta}{2}} |0\rangle
+ e^{i\phi} \cos{\frac{\theta}{2}} |1\rangle$.
Then, for any $\phi$, one can construct a corresponding unitary matrix
$U_G^{\phi} =
\left[\begin{array}{cr}
p & -q e^{-i\phi} \\
q e^{i\phi} & p \\
\end{array}\right]$,
such that
$U_G^{\phi} |\psi\rangle =
p |\psi\rangle +
q |\psi_\perp\rangle$ and
$U_G^{\phi} |\psi_\perp\rangle =
p |\psi_\perp\rangle - q |\psi\rangle$.
In the limit when $\phi = 0$, we recover the polar great
circle case
since $U_G^0 = U_P$. Thus if partial information $(\phi)$ is known,
given any arbitrary state, it is possible
to design a generalized Hadamard type gate for unequal superposition. Note
that for $p = q = \frac{1}{\sqrt{2}}$, $U_G^{\phi} = H_G^{\phi}$,
thereby yielding the result of Theorem~\ref{th2}.

\subsection{Unequal Equatorial Type Transformation}

The generalized version of the phase dependent Hadamard type of
transformation can be written as
\begin{equation}
\label{equ1x}
U(|\psi\rangle) = p|\psi\rangle + iq|\psi_{\perp}\rangle,
U(|\psi_{\perp}\rangle) = iq^*|\psi\rangle + p^*|\psi_{\perp}\rangle.
\end{equation}
Here again $p, q$ are known complex numbers with $|p|^2 + |q|^2 =
1$. Under the assumption that $\{|\psi\rangle,
|\psi_{\perp}\rangle\}$ transform in the same way as $\{|0\rangle,
|1\rangle\}$, $U$ is fixed to be $\left[\begin{array}{cr}
p & iq^* \\
iq & p^* \\
\end{array}\right]$.
In order to obtain classes of states obeying this transformation under
the action of $U$,
take $|\psi\rangle = a|0\rangle + b |1\rangle$ and
$|\psi_{\perp}\rangle = b^* |0\rangle - a^* |1\rangle$.
Ideally we should have
\begin{eqnarray}
U(|\psi\rangle) &=& p|\psi\rangle + iq|\psi_{\perp}\rangle
\nonumber \\
\ &=& p(a |0\rangle + b |1\rangle) +
        iq(b^* |0\rangle - a^* |1\rangle) \nonumber \\
\ &=& (ap+iqb^*) |0\rangle + (bp - iqa^*) |1\rangle. \nonumber
\end{eqnarray}
Then from linearity we get
\begin{eqnarray}
U(|\psi\rangle) &=& a U(|0\rangle) + b U(|1\rangle) \nonumber \\
\ &=& a (p|0\rangle + iq|1\rangle) +
      b (iq^*|0\rangle + p^*|1\rangle) \nonumber \\
\ &=& (ap+ibq^*) |0\rangle + (iaq+bp^*) |1\rangle \nonumber
\end{eqnarray}
Hence $pa + iqb^* = pa + iq^*b$, i.e., $qb^* = q^*b = (qb^*)^*$.
Thus $qb^*$ is real, i.e.,
\begin{enumerate}
\item both $q, b$ are real or
\item both $q, b$ are imaginary or
\item both $q, b$ are complex, with the constraint $\frac{b_1}{b_2} =
\frac{q_1}{q_2}$.
\end{enumerate}
Further, $pb - iqa^* = iqa + p^*b$, i.e., $iq(a + a^*) = b(p - p^*)$,
so $a_1 = \frac{b}{q} \cdot p_2$.

Hence we get a general class of states,
$|\psi\rangle = (\frac{b}{q} p_2 + i \cdot a_2) |0\rangle
+ b |1\rangle$ and $|\psi_\perp\rangle = b^* |0\rangle -
(\frac{b}{q} \cdot p_2 - i \cdot a_2) |1\rangle)$.
Depending on the above three possible solutions, i.e.,
whether $q, b$ are both real, or imaginary or
complex, we get different classes of states for which the unequal
superposition transformation (\ref{equ1x}) holds.
Again, it is easy to see that for the special case
$p = q = \frac{1}{\sqrt{2}}$, this reduces to the
class of states obtained in Theorem~\ref{th3}.

\section{V. Summary and Conclusions}
In this paper we have found the most general class of qubits (up to
isomorphisms) for which the Hadamard gate can be designed. This was
achieved using one of the fundamental axioms of quantum mechanics,
namely linearity. If expressed in the computational basis, the qubit
state assumes a specific form: one complex and one pure real
(imaginary) amplitude; the real (imaginary) parts of which are
equal. The Hadamard gate is universal for this class of ensemble,
i.e., it works for any state belonging to this particular ensemble.
When represented on a Bloch sphere, these states give a new
trajectory.Interestingly, it has some intersection points with the
polar and equatorial great circles.

Equal superposition of $|0\rangle$ and $|1\rangle$ states has played a
very crucial role in quantum algorithms to study various
problems, e.g., distinguishing between constant and balanced Boolean
functions~\cite{qDJ92}, database search~\cite{qGR96} etc. It would be
interesting to construct specific computational problems, which one can study
by exploiting the superposition of $|\psi\rangle$ and $|\psi_\perp\rangle$
from the class of states obtained in this paper.

We have also considered some Hadamard type transformations which
hold for polar and equatorial qubits, and have obtained some new
results. The situation becomes more general when the superposition
of the two amplitudes is not equal. Many new classes of states have
been found and all the results of the equal superposition case can
be recovered by letting the parameters to be equal.

The next step would be to generalize these results to higher
dimensions,where the analogue of the Hadamard transform would be the
discrete Fourier transform. In this direction we have obtained
partial results so far and further work is in progress.

\ \\
\noindent{\bf Acknowledgments:} We thank G. Kar and P. Mukhopadhyay
for useful discussions. PP acknowledges financial assistance from DST
under the SERC Fast Track Proposal scheme for young scientists.

\end{document}